\documentclass[oldversion]{aa} 
\usepackage{natbib}
\usepackage{graphicx}
\usepackage{txfonts}
\usepackage{calc}
\usepackage{color}
\usepackage{rotating}
\usepackage[USenglish]{babel} 

\begin{document}

\newcommand{\hinode}{\textsc{Hinode}}
\newcommand{\sunrise}{\textsc{Sunrise}}
\newcommand{\solarC}{\textsc{Solar-C}}
\newcommand{\carcsec}{$\mbox{.\hspace{-0.5ex}}^{\prime\prime}$}

\title{Vertical flows and mass flux balance of sunspot umbral dots}

\author{T.~L. Riethm\"uller\inst{1,2}
       \and
       S.~K. Solanki\inst{1,3}
       \and
       M. van Noort\inst{1}
       \and
       S.~K. Tiwari\inst{1}
       }

\institute{Max-Planck-Institut f\"ur Sonnensystemforschung (MPS),
           Max-Planck-Str. 2, 37191 Katlenburg-Lindau, Germany
     \and
           Technische Universit\"at Braunschweig, Institut f\"ur Geophysik und Extraterrestrische Physik,
           Mendelssohnstr. 3, 38106 Braunschweig, Germany
     \and
           School of Space Research, Kyung Hee University,
           Yongin, Gyeonggi, 446-701, Republic of Korea\\
           \email{[riethmueller;solanki;vannoort;tiwari]@mps.mpg.de}
          }

\date{Received; accepted}

%
\abstract
{
A new Stokes inversion technique that greatly reduces the effect of the spatial
point spread function of the telescope is used to constrain the physical properties
of umbral dots (UDs). The depth-dependent inversion of the Stokes parameters
from a sunspot umbra recorded with \hinode{} SOT/SP revealed significant temperature
enhancements and magnetic field weakenings in the core of the UDs in deep
photospheric layers. Additionally, we found upflows of around $960$~m/s in
peripheral UDs (i.e., UDs close to the penumbra) and $\approx 600$~m/s in
central UDs. For the first time, we also detected systematic downflows
for distances larger than 200~km from the UD center that balance the upflowing mass
flux. In the upper photosphere, we found almost no difference between the UDs
and their diffuse umbral background.

}

\maketitle
%

\section{Introduction}

Umbral dots (UDs) are small brightness enhancements in sunspot umbrae
or pores and were first detected by \citet{Chevalier1916}. The strong
vertical magnetic field in umbrae suppresses the energy transport by convection
\citep{Biermann1941}, but some form of remaining heat transport is needed to
explain the observed umbral brightness \citep{Adjabshirzadeh1983}. Magnetoconvection
in umbral fine structure, such as UDs and light bridges, is thought to be the main
contributor to the energy transport in the umbra \citep{Weiss2002}, see
reviews by \citet{Solanki2003}, \citet{Sobotka2006}, and \citet{Borrero2011}.

Progress in the physical understanding of umbral dots was made with
numerical simulations of 3D radiative magnetoconvection
\citep{Schuessler2006,Bharti2010}. Most of the simulated UDs have a horizontally
elongated shape and show a central dark lane in their bolometric intensity images.
In the deepest photospheric layers, the inner parts of UDs exhibit magnetic-field
weakenings and upflow velocities. The simulated UDs are surrounded by downflows that
are often concentrated in narrow downflow channels at the endpoints of the dark
lanes \citep{Schuessler2006}. Higher up in the photosphere, the UDs in the simulations
do not differ significantly from the diffuse background.

Considerable efforts on the observational side were made to test these theoretical
predictions. Dark lanes inside UDs were found in the observations of \citet{Bharti2007}
with the 50-cm \hinode{} telescope and by \citet{Rimmele2008}, who observed with the
76-cm Dunn Solar Telescope. However, \citet{Louis2012} analyzed straylight-corrected
\hinode{}/BFI data and did not find dark lanes in their observed UDs, which leaves
room for doubt whether the observed phenomena are really identical with the synthetic ones.
The UDs described in \citet{Bharti2007} differ from those reported in \citet{Schuessler2006}
in that the area of the observed features is an order of magnitude larger; possibly they
are the remains of a decayed light bridge.

More important than the dark lanes are the flows, since they are central to the convective
nature of the UDs. \citet{Riethmueller2008a}, using inversions of \hinode{}/SP data,
discovered upflows in the deep layers of peripheral UDs (PUDs) but not in central UDs (CUDs),
while downflows were not detected. Subsequently, \citet{Ortiz2010} studied a small pore recorded
with the CRISP instrument of the 1-m Swedish Solar Telescope and found irregular
and diffuse downflows in the range 500-1000~m/s for a small set of five UDs. In contrast,
in their recent study, \citet{Watanabe2012} analyzed a larger set of 339 UDs, also
observed with CRISP, and found significant UD upflows, but no systematic downflow
signals. Thus, the existence of downflows in or around UDs remains uncertain, so that
the fate of the material flowing up in UDs is unclear. The depth-dependent inversions
of full Stokes profiles derived in \citet{SocasNavarro2004} and later at higher resolution
in \citet{Riethmueller2008a} revealed a temperature enhancement and a field weakening
for the UDs compared to the nearby umbral background, which both were strongest in the
deepest observed layers.

Since the observational picture is inhomogeneous, there is a need for a more detailed UD study
for which high spatial and spectral resolution is of utmost importance. In this work, the improved
Stokes inversion method of \citet{vanNoort2012} is applied to \hinode{}/SP data
\citep[see][]{vanNoort2013}. This so-called 2D inversion method allows the depth-dependent
structure to be obtained basically as it would be in the absence of the telescope's
point spread function (PSF).


\section{Observation, data reduction, and analysis}\label{SecObs}

\begin{figure}
\centering
\includegraphics[width=80mm]{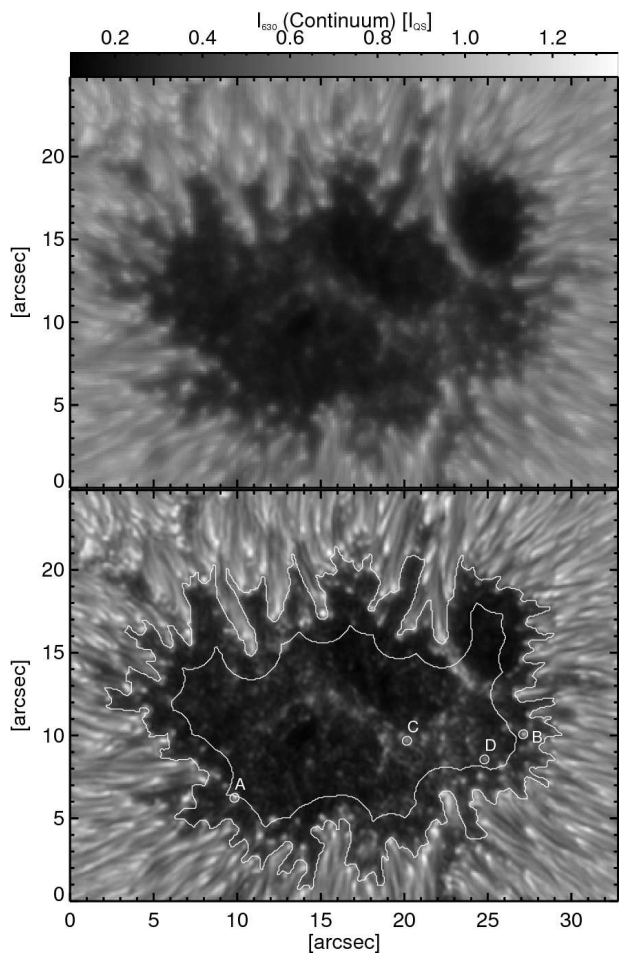}
\caption{Stokes~$I$ continuum intensity of the \hinode{}/SP map of a sunspot umbra of NOAA AR~10933.
The original data are plotted in the top panel. The Stokes~$I$ continuum resulting from the 2D inversion is shown in
the bottom panel. The intensity is normalized to the mean quiet-Sun intensity $I_{\rm{QS}}$. The
outer contour line in the bottom panel indicates the edge of the umbra as retrieved from the magnetic field inclination map (see
main text), the inner contour line separates central from peripheral umbral dots (UDs).
Four typical UDs are marked by circles and letters.}
\label{Fig1}
\end{figure}

The data we analyzed in this study were recorded from 12:43 to 13:00~UT on 2007 January 5
with the spectropolarimeter \citep[SP,][]{Lites2001} of the Solar Optical Telescope
\citep[SOT,][]{Tsuneta2008} on the \hinode{} spacecraft \citep{Kosugi2007}.
The SP was operated in its normal map mode, i.e., the integration time per slit position
was 4.8~s, resulting in a noise level of $10^{-3}$ (in units of the continuum intensity).
The sampling along the slit, the slit width, and the scanning step size were 0\carcsec{}16,
the spectral sampling in the considered range from 6300.89 to 6303.26~{\AA} was 21~{m\AA}~pixel$^{-1}$.
The center of the observed umbra was located very close to the disk center, at a heliocentric angle
of 2.6$^\circ$. The full Stokes profiles were corrected for dark current as well as flat-field effects
and calibrated with the $sp\_prep$ routine of the SolarSoft package. Part of the calibrated Stokes~$I$
continuum intensity map obtained with \hinode{} SP is shown in the upper panel of Fig.~\ref{Fig1}. The original FOV
is much larger and contains quiet-Sun regions that are used for the intensity normalization.

Under the assumption of local thermodynamic equilibrium, the Stokes profiles of the Fe\,{\sc i}
6301.5~{\AA} and 6302.5~{\AA} lines were inverted by applying the version of the SPINOR inversion
code \citep{Frutiger2000a,Frutiger2000b} extended by \citet{vanNoort2012}. In this version of the code,
the instrumental effects responsible for the spectral and spatial degradation of the observational data
are taken into account, so that the inverted parameters correspond to spatially deconvolved
values (but without the added noise that deconvolution generally introduces). The observational data are
spatially upsampled by a factor of two, so that the input and output data of the SPINOR inversion have
a sampling of 0\carcsec{}08 per pixel \citep[for details, see][]{vanNoort2013}. According to
\citet{vanNoort2012}, the spatial PSF used for the inversion considers the 0.5~m clear aperture of the SOT,
the central obscuration, the spider \citep{Tsuneta2008,Suematsu2008}, and a defocus of 0.1~waves,
even though the focus position of our data set was not accurately known.
Height-dependent temperature, LOS velocity, magnetic field strength, field inclination, field azimuth,
and micro-turbulence are determined at three $\log{\tau_{500}}$ nodes: $-2.5$, $-0.9$, and $0$.
More details of the inversion of this spot are provided in \citet{vanNoort2013} and \citet{Tiwari2013}.

A continuum map obtained from the best-fit Stokes~$I$ profiles of the 2D inversion result can be seen
in the bottom panel of Fig.~\ref{Fig1}. Since the deconvolution of the data with the theoretical spatial
PSF is now indirectly part of the inversion process, the contrast is significantly enhanced and
UDs can be identified much more clearly than in the original data. Hence, the continuum map
in the bottom panel of Fig.~\ref{Fig1} was used for a manual detection of the location of the most
prominent 67 UDs. Misidentification of brightness features that are separated from
the penumbra in the continuum image but are still connected to the penumbra in the magnetic field inclination
map were excluded since we defined the boundary of the umbra by thresholding the lowpass-filtered inclination
map (7x7 pixels) at 40 degrees and corrected the boundary found in this way manually in a few doubtful
cases where the penetration of long and narrow penumbral filaments led to a wrong result. The umbral
boundary is shown as the outer contour line in the bottom panel of Fig.~\ref{Fig1}. Furthermore, we divided
the set of 67 UDs into 23 central UDs (CUDs) and 44 peripheral UDs (PUDs). The inner contour line in the
bottom panel of Fig.~\ref{Fig1} separates the PUDs from the CUDs. The criterion used is simply the
distance to the umbra-penumbra boundary. CUDs are $>$\,1000\,km away from this boundary, PUDs $\le$\,1000\,km.

Once the locations of the UDs' centers were known, the UD boundaries were determined from the continuum map
by a multilevel tracking (MLT) algorithm \citep[see][]{Bovelet2001}, using 25 equidistant intensity levels.
The resulting contiguous MLT structures were then cut at 50\% of the local min-max intensity range, which was
taken as the UD boundary. A detailed description of the use of the MLT algorithm for isolating UDs is given
in \citet{Riethmueller2008b}. 

The knowledge of the UD boundaries allowed us to average UD properties over all pixels within the
UD boundary. Stratifications of temperature, LOS velocity, and field strength of the UDs were then determined
as such averages. The same quantities were also determined for the UDs' diffuse background (DB), defined
as the average over all pixels (ignoring penumbral pixels) in a 400-km-wide ring around the
UD boundary. UD and DB quantities were retrieved for optical depths between $\log{\tau_{500}}=-2.5$
and 0 in steps of $0.5$.

The LOS velocity maps of the inversion result show a clear $p$-mode pattern with a spatial wavelength of
about 10\arcsec{} that has to be removed to avoid any $p$-mode influence on our results. The usually employed
technique of Fourier filtering in 3D $k\omega$-space cannot be applied in our case because
only a single map of the observational data was available for inversion. We therefore removed the
$p$-modes in the LOS velocity maps at all used $\log{\tau_{500}}$ nodes by applying a highpass filter
(implemented as the difference between the original velocity map and its running boxcar, $21\times{}21$ pixels,
filtered counterpart). Since our results depend on a careful zero-velocity determination, we re-calibrated
the velocities even if the highpass filter already roughly removed the velocity offset. To achieve this
we assumed that the dark core of the umbra is at rest. The darker part of the umbra is identified by
thresholding the lowpass-filtered continuum image ($11\times{}11$ pixels) at 50\% of the intensity range.
We furthermore excluded a circle of 500~km radius around each of the 67 identified UDs and subtracted the
mean velocity of the remaining dark umbral pixels from the velocity maps at each optical depth. This
procedure was found to be robust in the sense that changing the threshold for identifying the darkest part
of the umbra by $\pm10\%$, or increasing the radius of the exclusion zone around the UDs by 200~km, did not
influence our results.

\section{Results}
Even at the significantly improved image quality provided by the inversion, we were unable
to find dark lanes in the central parts of the UDs as reported in \citet{Schuessler2006} in
MHD simulations and in \citet{Bharti2007} in other deconvolved \hinode{} images.

\begin{figure}
\centering
\includegraphics[width=\linewidth]{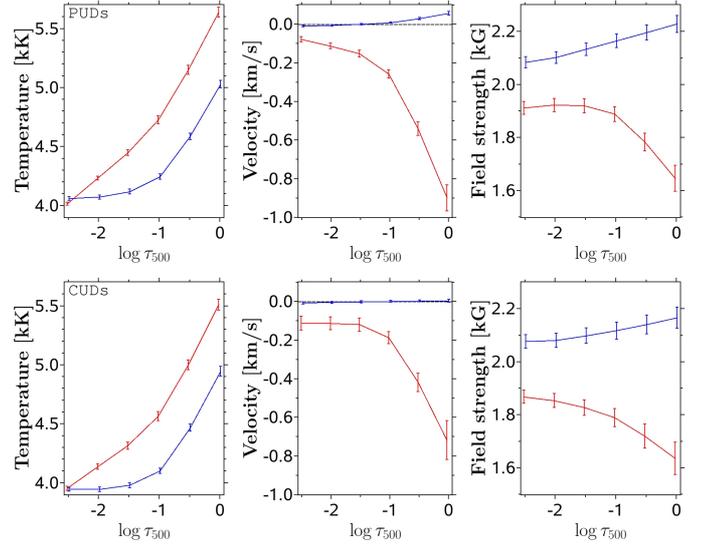}
\caption{Optical-depth dependence of temperature, LOS velocity, and magnetic field strength averaged
over 44 peripheral umbral dots (top three panels) and 23 central umbral dots
(bottom three panels). The error bars denote standard deviations of the mean ($\sigma/\sqrt{N}$). The red
lines exhibit the stratifications of the mean UD, while the blue lines correspond to the mean nearby
diffuse background (see main text for details). Negative LOS velocity values indicate upflows.}
\label{Fig2}
\end{figure}

The stratifications of temperature, velocity, and field strength, averaged separately over all PUDs
and CUDs, are displayed in Fig.~\ref{Fig2}. While in the upper photosphere ($\log{\tau_{500}}=-2.5$)
the considered properties hardly differ between the mean UD and DB, they deviate significantly from
each other in the deep photosphere ($\log{\tau_{500}}=0$). Compared with their DB, we find a
temperature enhancement and field-weakening of 610~K and 580~G at optical-depth unity for the
mean PUD, while for CUD the values are 570~K and 530~G. The mean UD magnetic field
weakens with depth, while the field strength of the mean DB increases with depth, as expected.

\begin{figure}
\centering
\includegraphics[width=\linewidth]{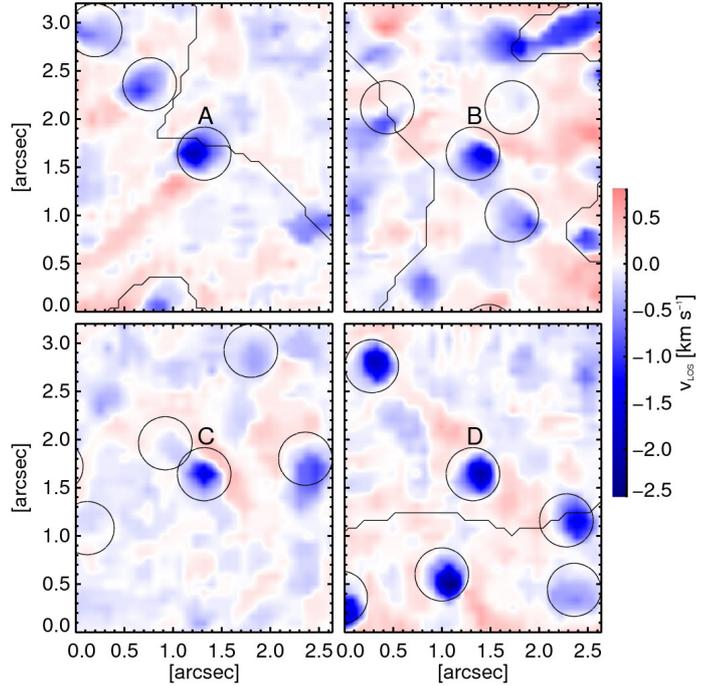}
\caption{LOS velocities at optical-depth unity of typical UDs marked as circles of 200~km
radius around the position of the UD's peak intensity. The UDs in the center of each panel identified
by letters are the same as in the bottom panel of Fig.~\ref{Fig1}.}
\label{Fig3}
\end{figure}

\begin{figure}
\centering
\includegraphics[width=\linewidth]{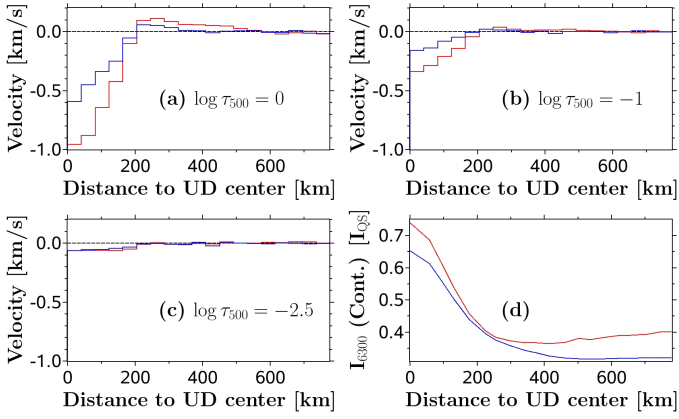}
\caption{Panel~(a)-(c): LOS velocity at constant optical depth as function of the distance to the UD center averaged
over all azimuthal angles of the peripheral (red lines) and central (blue lines) UDs. The optical depths
are given as text labels. Panel~(d): Mean continuum intensity profile of the two UD classes.}
\label{Fig4}
\end{figure}

The LOS velocity (which is virtually identical to the vertical velocity component due to the small
heliocentric angle) of DB and UD is almost zero in the upper photosphere. Strong upflows of
$-900$~m/s and $-720$~m/s are found at optical-depth unity for the mean PUD and CUD, while a weak
but significant downflow of 57~m/s is found for the mean DB of the peripheral UDs only. These values
should be compared with the uncertainty in the velocity averaged over the DB of PUDs,
$\sigma/\sqrt{N}=10$~m/s ($\sigma$ - standard deviation of the 44 DB velocities from
their mean value, $N=44$ - number of UDs). The DB of the central UDs is on average at rest within the
error bars. A better insight into the up- and downflows in and around UDs is given in Figs.~\ref{Fig3}
and \ref{Fig4}. The LOS velocity maps at $\log{\tau_{500}}=0$ are shown in Fig.~\ref{Fig3} for four typical UDs
and reveal that the upflows are mainly concentrated within the marked 200~km vicinity of the UD center, while
weaker downflows are preferentially found farther out. In general, the downflows only partly surround a UD,
they are often concentrated on one or two sides of the UD. Although weak upflows are also present in the
vicinities of UDs, the downflows dominate for the PUDs, as deduced from Figs.~\ref{Fig2} and ~\ref{Fig4}.
For a better visibility the spatial sampling of the maps was increased via bi-linear interpolation.

In Fig.~\ref{Fig4}, we averaged the velocities of all pixels within rings of 40~km width around 
the UD center (ignoring penumbral pixels) and plotted the mean velocities as a function of the distance to the
UD's center, again separately for PUDs and CUDs. Note that this method is independent of any determination
of the UD boundary. Panel~(a) of Fig.~\ref{Fig4} shows the velocities at optical-depth unity. Between $0-200$~km
distance from the UD's center, we find upflows (see enlarged velocity maps in Fig.~\ref{Fig3}
where the UDs are marked with circles having a 200~km radius). Then, from 200 to 500~km we see downflows (between 200
and 350~km for CUDs), while for distances larger than 500~km the velocity is almost zero. The downflows
in the lower photosphere peak at a distance of roughly 240~km from the UD's center and have values of
110~m/s and 58~m/s for the mean PUD and CUD. They are thus minute compared to the maximum
upflows in the UDs of $-960$~m/s and $-600$~m/s. The upflows and downflows
are on average stronger for the PUDs than for the CUDs. Both up- and downflows increase rapidly with depth
(compare panels (a)-(c)). The upflows within the UDs are much weaker at $\log{\tau_{500}}=-1$
(panel~(b)) and the downflows cannot be seen anymore. At $\log{\tau_{500}}=-2.5$ (panel (c)), there
is almost no velocity signal. The mean intensity profiles are plotted in panel~(d) and reveal
a half-width-half-maximum radius of 120~km for the PUDs and 140~km for the CUDs.

We next calculated mass fluxes as the sum over all pixels within a 500-km vicinity of the UD center since
for larger distances the velocity is negligible. The required densities were provided by the SPINOR code
under the assumption of hydrostatic equilibrium. Table~\ref{MassFluxes} lists the upward and downward
mass fluxes per UD, $M_u$, and $M_d$ for various optical depths and separately for the two UD classes.
The uncertainties in Table~\ref{MassFluxes} are the standard deviations of the averages over all UDs
of a given class ($\sigma$). The last two columns give the mass flux ratios. The mass flux increases
strongly with depth due to the density and velocity increase. Even if the method of azimuthal
averaging leads to downflow velocities that are lower than the upflow velocities, the averaging is
performed over a much larger area so that the upward and downward mass flux are roughly balanced within the
uncertainties.
\begin{table*}
\caption{Upward and downward mass fluxes per umbral dot (computed within 500~km radii) and their ratios at various optical depths.}
\label{MassFluxes}                                
\centering                                        
\begin{tabular}{c c c c c c c}                    
\hline                                            
\noalign{\smallskip}
$\log{\tau_{500}}$ & $M_u$                             & $M_u$                             & $M_d$                             & $M_d$                             & $M_u/M_d$                        & $M_u/M_d$                        \\
                   & PUD                               & CUD                               & PUD                               & CUD                               & PUD                              & CUD                              \\
                   & [\tt{1O}$^8$\,kg/s] & [\tt{1O}$^8$\,kg/s] & [\tt{1O}$^8$\,kg/s] & [\tt{1O}$^8$\,kg/s] & []                               & []                               \\
\hline                                                                                                                                       
\noalign{\smallskip}                                                                                                                         
\tt{~O~~}          & \tt{6OO$\pm$13O}    & \tt{55O$\pm$12O}    & \tt{67O$\pm$17O}    & \tt{5OO$\pm$~82}    & \tt{O.89$\pm$O.42} & \tt{1.O9$\pm$O.41} \\
\tt{-1~~}          & \tt{13O$\pm$~37}    & \tt{15O$\pm$~32}    & \tt{14O$\pm$~36}    & \tt{14O$\pm$~26}    & \tt{O.99$\pm$O.53} & \tt{1.O3$\pm$O.4O} \\
\tt{-2~~}          & \tt{~37$\pm$~12}    & \tt{~41$\pm$~11}    & \tt{~33$\pm$~12}    & \tt{~36$\pm$~~7}    & \tt{1.12$\pm$O.74} & \tt{1.13$\pm$O.52} \\
\tt{-2.5}          & \tt{~2O$\pm$~~7}    & \tt{~22$\pm$~~6}    & \tt{~18$\pm$~~7}    & \tt{~19$\pm$~~4}    & \tt{1.16$\pm$O.83} & \tt{1.18$\pm$O.6O} \\
\hline                                            
\end{tabular}
\end{table*}

%

\section{Discussion and conclusions}
We have used a new inversion technique to retrieve the atmospheric parameters of 67 UDs
in a sunspot umbra from data recorded with the spectropolarimeter onboard \hinode{}.
In agreement with earlier studies \citep{SocasNavarro2004,Riethmueller2008a}, we found
that in the deep photosphere the temperature is enhanced and the magnetic field is weakened
in the UDs compared with their umbral surroundings. Table~\ref{LitComp} compares the main
UD properties retrieved from the conventional and the improved inversion technique. For
a direct comparison with \citet{Riethmueller2008a}, who reported peak values and not
spatial averages, we also listed the peak values obtained from the new inversion in
Table~\ref{LitComp}. In fact, all values listed in Table~\ref{LitComp} are higher
for the new inversion method, which emphasizes the considerably improved data quality reached by
the implicit removal of the telescope's spatial PSF.

\begin{table}
\caption{Comparison of UD properties at the continuum formation height between \citet{Riethmueller2008a} and this study.}
\label{LitComp}                                   
\centering                                        
\begin{tabular}{c c c c c c c}                    
\hline                                            
\noalign{\smallskip}
UD class                            & PUD        & PUD                     & PUD                    & CUD        & CUD                     & CUD                    \\
study                               & \tt{2OO8}a & this                    & this                   & \tt{2OO8}a & this                    & this                   \\
value type                          & peak       & peak                    & avg                    & peak       & peak                    & avg                    \\
\hline
\noalign{\smallskip}
$T_{\rm{UD}}-T_{\rm{DB}}\,$[\tt{K}] & \tt{57O}   & \tt{~91O} & \tt{61O} & \tt{55O}   & \tt{~83O} & \tt{57O} \\
$B_{\rm{DB}}-B_{\rm{UD}}\,$[\tt{G}] & \tt{51O}   & \tt{~7OO} & \tt{58O} & \tt{48O}   & \tt{~84O} & \tt{53O} \\
$v_{\rm{up}}\,$[\tt{m/s}]           & \tt{8OO}   & \tt{173O} & \tt{96O} & \tt{-~~}   & \tt{129O} & \tt{6OO} \\
$v_{\rm{down}}\,$[\tt{m/s}]         & \tt{-~~}   & \tt{-~~~} & \tt{11O} & \tt{-~~}   & \tt{-~~~} & \tt{~58} \\
\hline                                            
\end{tabular}
\end{table}

The 2D inversion results revealed clear upflow signals for both UD types, while \citet{Riethmueller2008a}
could only find them for the PUDs. On average, UDs show upflows up to a radial distance of 200~km from
their centers. In general, these upflows are stronger for PUDs than for CUDs. Between 200 and 500~km
from a UD's center, we found low but significant downflows, whereas there is no relevant velocity signal
farther away. The velocity signal decreases rapidly with atmospheric height.

Previous observational studies detected upflows, but could not detect downflows associated
with UDs \citep[e.g.][]{SocasNavarro2004,Riethmueller2008a,Watanabe2012}, or at least not systematically
\citep{Ortiz2010}. This raised the question where all the upflowing plasma ends up. The first systematic
detection of downflows around UDs in this paper gives us the possibility of calculating upward and
downward mass fluxes. Our finding of the very well-balanced mass fluxes depends on the careful velocity calibration
described in Sect.~\ref{SecObs}. If all umbral pixels had been used for the zero velocity determination,
the zero velocity could possibly be blueshifted due to the UD upflows, thus giving rise to artificial
downflows. This effect is ruled out since our velocity re-calibration ignores the UDs and uses the darkest
parts of the umbra only.

We furthermore believe that the downflows seen in the top left panel of Fig.~\ref{Fig4} are real and not a result
of ringing effects. Such effects could be caused by the nearly axisymmetrical shape of the spatial PSF used
in our inversion and would have affected all quantities. However, plots of temperature and field strength versus
the UD center distance (not shown) do not exhibit any signs of ringing. According to \citet{Schuessler2006}, the
downflows are concentrated in narrow channels preferentially at the endpoints of the central dark lanes
of the UDs. Our spatial resolution is insufficient to detect the dark lanes or the narrow downflow
channels. The relatively weak downflow signals become significantly stronger than the noise only after
the azimuthal averaging around UDs.

The picture introduced in \citet{Schuessler2006} of UDs as a natural consequence of magnetoconvection
in the strong vertical magnetic field of an umbra is qualitatively confirmed by our study.
In deep layers the rising hot plasma pushes the field to the side, weakening the field there. Around the
continuum formation height the rising gas cools through radiative losses, turns over, and flows down around
the UDs. This evidence for overturning comes from the mass balance between up- and downflows and from the fact
that the central upflows are associated with hot material, whereas the peripheral downflows are cool.
Furthermore, the rapid decrease of upward mass flux with height is also typical for overturning, overshooting
convection. The fact that PUDs are associated with higher flow velocities and slightly higher temperature
enhancements suggests a more vigorous convection than in the central umbra. We note, however,
that our results differ in one detail from the simulations. Whereas the simulations produce concentrated
downflows on opposite ends of the UDs, the observations reveal a more diffuse downflow structure.

The 2D inversion technique also returns an enhanced temperature excess and magnetic-field reduction compared
with traditional inversions. The temperature and magnetic field anomalies turn out to be very
similar for PUDs and CUDs (within 10\% of each other).

We suggest observations at a spatial resolution higher than available here, e.g., with the re-flight of
the 1~m \sunrise{} telescope \citep{Solanki2010,Barthol2011}, or with the envisaged 1.5-m \solarC{} telescope,
to determine if the UDs show the predicted central dark lanes with narrow downflow channels at their
endpoints.

\begin{acknowledgements}
\hinode{} is a Japanese mission developed and launched by ISAS/JAXA, with NAOJ, NASA, and STFC (UK)
as partners. This work has been partly supported by the WCU grant No. R31-10016 funded by the Korean
Ministry of Education, Science \& Technology.
\end{acknowledgements}


\end{document}